\documentclass[11pt,a4paper]{article}
\usepackage{amsfonts,amscd,amsmath,amssymb,amsthm}
\newtheorem{lem}{Lemma}[section]
\newtheorem{theo}{Theorem}[section]

\newtheorem{ex}{Example}[section]

\DeclareMathOperator{\tr}{Tr}

\newcommand{\1}{\mathbb{I}}

\title{On a quantum version of Shannon's conditional entropy}
\author{ R. Schrader\thanks{e-mail: schrader@physik.fu-berlin.de, 
Supported in part by
DFG SFB 288 ``Differentialgeometrie und Quantenphysik''}\\
Institut f\"{u}r Theoretische Physik\\
Freie Universit\"{a}t Berlin, Arnimallee 14\\
D-14195 Berlin, Germany}

\begin{document}
\maketitle

\begin{abstract}
In this article we propose a quantum version of Shannon's
conditional entropy. Given two density matrices $\rho$ and $\sigma$ on
a finite dimensional Hilbert space and with $S(\rho)=-\tr\rho\ln\rho$ being 
the usual von Neumann entropy, this quantity $S(\rho|\sigma)$ is
concave in $\rho$ and satisfies $0\le S(\rho|\sigma)\le S(\rho)$, a 
quantum analogue of Shannon's famous inequality. Thus we view 
$S(\rho|\sigma)$ as the
entropy of $\rho$ conditioned by $\sigma$. The second inequality is 
an equality if $\sigma$ is a multiple of the identity. In contrast to the
classical case, however, $S(\rho|\rho)=0$ if and only if the
non-vanishing eigenvalues of $\rho$ are all non-degenerate. Also in
general and again in contrast to the corresponding classical situation 
$S(\rho,\sigma)=S(\sigma)+S(\rho|\sigma)$ is not symmetric
in $\rho$ and $\sigma$ even if they commute. We also show that there
is no quantum version of conditional entropy in terms of two 
density matrices, which shares more properties with the classical case
and which in particular reduces to the classical case when the two
density matrices commute. As an alternative we propose to use spectral
resolutions of the unit matrix instead of density matrices. We briefly
compare this with the algebraic approach of Connes and St{\o}rmer and Connes,
Narnhofer and Thirring.  
\end{abstract}

\section{Introduction}
 The concept of entropy plays a
 major role in thermodynamics and statistical mechanics. It serves to 
describe the behavior of macroscopic systems. The name ``entropy'' was
introduced by Clausius (1865) and derives from  
$\epsilon\nu\tau\rho o\pi\iota\eta$ ``transformation''. 
It was von Neumann (1927 \cite{vN2}), who generalized the 
classical expression of Boltzmann and Gibbs for the entropy to 
quantum mechanics by using the concept of what is now called a density 
matrix, also introduced quite generally by him in the same year \cite{vN1}. 
In the special context of radiation damping the density matrix was
 discovered independently by L. Landau \cite{L} and by F. Bloch 
\cite{Bloch}, again in the 
same year (see also the citation in \cite{LL}). For
a technical overview of the developments up to 1978 and with further 
historical references see \cite{Wehrl}. For recent expositions 
see \cite{OP,Nielsen}. In the theory of dynamical systems 
entropy and the derived notion of topological entropy also plays an
 important role, see e.g. the contributions in \cite{Sinai}. 

In a seminal article Shannon (1948, \cite{Shannon})
introduced the concept of entropy into information theory. Roughly
 speaking a gain in information means a decrease in entropy. Shannon
 also provided the concept of conditional 
entropy. It is a measure how entropy is reduced given a preexisting knowledge. 
To the author's best knowledge the first construction in quantum
mechanics coming close to such a notion is due to E. Lieb \cite{Lieb}
(see also \cite{Wehrl,Nielsen}). It involves tensor
 product structures and it was called a relative entropy in
 \cite{Lieb} (but a conditional entropy in \cite{Wehrl}, p. 259).
 In view of recent developments in quantum computation
 and quantum coding (see \cite{Preskill,Nielsen} for a concise account) it
 is highly desirable to have such a quantity at ones disposal. 
 There is a construction of a non-commutative analogue of Shannon's
 conditional entropy by Connes and St{\o}rmer \cite{CS} and Connes,
 Narnhofer and Thirring \cite{CNT}(for an exposition and a discussion
 of further 
 developments see e.g. \cite{Be,OP}). 
 More recently attempts have been made to construct a mutual 
 information analogous to Shannon's conditional entropy in the context
 of quantum error-correction.  In two of these attempts \cite{Schu,Lloyd}, made
 independently, yielded the same quantity. The first article exhibits 
  necessary and sufficient conditions for quantum
 error-correction to be possible in terms of the mutual information
 like the
quantity given there, and a conjecture is made on its connection with quantum
channel capacity, explored in more detail \cite{Ba}. The connection
with channel capacity was also analyzed in \cite{Lloyd}. In 
\cite{Nielsen2} its connection with entanglement is discussed. 
In yet another approach \cite{Le} the starting point is one density
matrix on a tensor product. The conditioning is then obtained by
looking at the two density matrices in the two sub-systems resulting
by taking the corresponding partial traces.

In this article we will propose a different candidate for a quantum mechanical
conditional entropy $S(\rho|\sigma)\ge 0$, a function of two
 density matrices $\rho$ and $\sigma$ in a same Hilbert space and 
having the interpretation of
the entropy of $\rho$ conditioned by the ``knowledge'' given by
 $\sigma$. For simplicity we will only discuss the finite dimensional
 case
although an extension to the infinite dimensional case seems possible.
If we view $\rho$ as the analogue of $X$ and $\sigma$
the analogue of $Y$ such that von Neumann's entropy $S(\rho)$ is the
analogue of Shannon's entropy $H(X)$, then       
this conditional entropy shares several but not all properties of
 Shannon's conditional entropy $H(X|Y)$ (see section 3 for a brief 
recapitulation of Shannon's theory).  
In particular the ``knowledge'' of
 $\sigma$ reduces the entropy, i.e. the inequality $S(\rho|\sigma)\le
 S(\rho)$ holds. This corresponds exactly to
 Shannon's famous inequality $H(X|Y)\le H(X)$ and was our main
 motivation for our construction. Also and again in analogy to the
 classical theory we wanted the conditioning to be given by a
 quantity on the same footing as the original density matrix,
 i.e. conditioning should also be given by a density matrix. If as in the
 classical case  
 $\sigma$ contains no information, i.e. if it is a multiple of the
 identity such that $S(\sigma)$ is maximal, then
 $S(\rho|\sigma)=S(\rho)$. In contrast to the classical
 case $H(X|X)=0$, however, the relation $S(\rho|\rho)=0$ holds
 if and only if the non-zero eigenvalues of $\rho$ are
 non-degenerate. In particular $S(\rho|\rho)=0$ if $\rho$ is pure. We
 will not elaborate on the question, whether the failure of our 
$S(\rho|\sigma)$ to satisfy all
 corresponding classical properties, like this last property, is due to a  
fundamental difference of quantum  and classical
 information theory. In particular we will not provide a more detailed
 quantum mechanical interpretation of
 $S(\rho|\sigma)$. Also so far we have not analyzed whether it may be
 used in the context of channel capacity. Rather we will argue that
other quantum mechanical versions of conditional entropy, which
 share more properties with the classical counterpart $H(X|Y)$, do not
 exist. 

The article is organized as follows. In section 2 we provide 
the construction of a quantum version $S(\rho|\sigma)$ of the
conditional entropy and establish several properties. In section 3 and
after a brief review of
Shannon's theory we compare this with Shannon's conditional 
entropy. In section 4 we first present a list of desirable properties
for a quantum version of conditional entropy given in terms of two
density matrices. We then show that even parts of these desiderata can
not be fulfilled simultaneously. In particular there is no version
involving two density matrices and which reduces to the classical
case, when these two density matrices commute. We will provide an 
alternative in terms of resolutions of the unit matrix in terms of 
orthogonal projections and which share more properties with the
classical case. Briefly we will compare this ansatz with the algebraic
constructions given by Connes and St{\o}rmer and Connes,
Narnhofer and Thirring.

\section{Construction of a quantum conditional entropy}

Let $\rho$ be a density matrix on a finite dimensional Hilbert space 
$\cal{H}$, i.e $\rho\ge 0$ and $\tr\rho=1$, where $\tr$ denotes the
canonical trace on $\cal{H}$. We write $\rho =\sum_i\rho_i\;P_i$ for
the spectral representation of $\rho$ where the projections $P_i\neq
0$ are pairwise orthogonal ( i.e. $P_{i}P_{j}=\delta_{ij} P_{i}, 
P_{i}=P_{i}^{\dagger}$), such that $\rho_i\ge 0$, $\rho_i\neq\rho_j$
for $i\neq j$ and $\sum_{i}P_{i}=\1$, where $\1$ is the identity
operator on $\cal{H}$. Thus $\tr\rho=\sum_i \dim P_i\;\rho_i=1$ with
$\dim P=\tr P=\dim P\cal{H}$
for any projection $P$. Here and in what follows projection operators are
always understood to be orthogonal. With this notational
convention the $P_i$ are canonically defined in terms of $\rho$. Since
this fact will be crucial in what follows, let us briefly recall a standard
proof. The eigenvalues $\rho_{i}$ (and their 
degeneracies $(=\dim\,P_{i})$) are of course uniquely determined by
$\rho$ as solutions
in $\lambda$ of the secular equation $\det(\lambda\1-\rho)=0$, a basis
independent relation, such
that $\det(\lambda\1-\rho)=\prod_{i}(\lambda-\rho_{i})^{\dim P_{i}}$. 
Order the $\rho_{i}$ in such a way that $1\ge\rho_1 >\rho_2 >\rho_3
>...$ . 
Then $P_{1}=\lim_{n\rightarrow\infty}(\rho/\rho_{1})^n$, 
$P_{2}=\lim_{n\rightarrow\infty}((\rho-\rho_{1}P_{1})/\rho_{2})^n$, etc.

The quantum mechanical entropy of $\rho$ is given as 
$S(\rho)=-\sum_i \dim P_i\;\rho_i\ln \rho_i$, which is continuous and
concave in $\rho$ (for an account of sub-additivity and convexity
properties of the entropy and related quantities see e.g. 
\cite{Lieb,Wehrl, Nielsen}). Let $\sigma$ be another density
matrix on the same space $\cal{H}$ with the spectral representation 
$\sigma =\sum_j \sigma_j Q_j$ again written in a canonical way.
We define the conditional entropy by
\begin{eqnarray}
\label{def1}
S(\rho|\sigma)&=&\sum_j \dim Q_j\;\sigma_j\;F(\rho,Q_j)\nonumber\\
              &=&\sum_{j}\tr Q_{j}\sigma\;F(\rho,Q_{j})
\end{eqnarray} 
where
\begin{equation}
\label{def2}
F(\rho,Q)= -\tr(Q\rho Q\ln(Q\rho Q))+\tr(Q\rho Q)\ln \tr(Q\rho Q)
\end{equation}
for any orthogonal projection $Q$. Since the $Q_{j}$'s and
$\sigma_{j}$'s are well defined
in terms of $\sigma$ and since trivially $Q\rho Q\ge 0$, 
$S(\rho|\sigma)$ is well defined.
Also as usual in this context $A\ln A$ for any non-negative operator
$A$ is defined in terms of the spectral representation of $A$ with 
the natural convention that $x\ln x|_{x=0}=0$.
If $Q\rho Q\neq 0$ then also $0\neq\tr Q\rho Q=\tr Q\rho$
and then we may write 
\begin{equation}
\label{F}
F(\rho,Q)=\tr Q\rho \cdot S(\rho_{Q})
\end{equation}
with
\begin{equation}
\rho_{Q}=\frac{1}{\tr(Q\rho)}
\cdot Q\rho Q
\end{equation}
being a density matrix. Actually we might use \eqref{F} instead of 
\eqref{def2} as a definition for $F(\rho,Q)$ with the convention,
usually made in similar contexts (see e.g. \cite{Top}), that 0 times 
something undefined is 0. Relation \eqref{F} shows that $F(\rho,Q)\ge
0$ for all $\rho$ and $Q$.
Using \eqref{F} we may rewrite $S(\rho|\sigma)$ as
\begin{equation}
\label{def3}
S(\rho|\sigma)=\sum_{j\;:\; Q_{j}\rho Q_{j}\neq 0}\tr Q_{j}\sigma\;
\tr Q_{j}\rho\; S(\rho_{Q_{j}}).
\end{equation}
There is yet another way of writing $F(\rho,Q)$. It uses the
relative entropy
$0\le\,S_{rel}(A,B)=\tr\,A(\ln\,A-\ln\,B)\le\,\infty$, 
which is defined for any $A\ge 0$ and $B\ge 0$. The relative entropy is lower
semi-continuous in $A$ and jointly convex in $A$ and $B$, see e.g. 
\cite{OP,Wehrl}. Obviously 
$S_{rel}(\lambda\,A,\lambda\,B)=\lambda\,S_{rel}(A,B)$ holds for any 
$\lambda>0$ and we have
\begin{equation}
\label{F2}
F(\rho,Q)=-S_{rel}(Q\,\rho\,Q,\tr(Q\,\rho\,Q)\1)
\end{equation}
such that 
\begin{equation}
\label{def4}
S(\rho|\sigma)=-\sum_{j}\tr Q_{j}\sigma \cdot 
S_{rel}(Q_{j}\rho Q_{j},\tr(Q_{j}\rho Q_{j})\1)
\end{equation} 
It is instructive to compare $S(\rho|\sigma)$ with
$S(E_{\underline{Q}}(\rho))$ and which actually motivated our
construction of $S(\rho|\sigma)$. $E_{\underline{Q}}$ is the linear map
on the set of linear operators $A$ on $\cal{H}$ given as  
$E_{\underline{Q}}(A)=\sum_{j}Q_{j}A Q_{j}$ . The $Q_{j}$'s are as 
above, i.e a any set $\underline{Q}=\{Q_{j}\}$
of pairwise orthogonal nonzero projection operators with $\sum_{j} Q_{j}=\1$
and which is called a resolution of the identity. $E_{\underline{Q}}$
  is a conditional expectation (see e.g. \cite{GHJ}) with range being the
  $\star$-algebra consisting of all linear operators which commute
  with all $Q_{i}$. In particular $E_{\underline{Q}}$ maps density
  matrices into density matrices. More precisely, let 
${\cal B}={\cal B}({\cal H})$ be the $\star$-algebra of all linear
operators on $ {\cal H}$, which is (isomorhic to) a full
matrix-algebra. Then $E_{\underline Q}({\cal B})$ is a 
$\star$-sub-algebra of ${\cal B}$ and the direct sum of the
$\star$-sub-algebras $Q_{j}{\cal B}Q_{j}={\cal B}(Q_{j}{\cal H})$, which
are (isomorphic to) full matrix algebras. Although any finite dimensional
$\star$-algebra is (isomorphic to) a direct sum of full matrix algebras,
not all $\star$-sub-algebras of ${\cal B}$ are of the form 
$E_{\underline{Q}}({\cal B})$ for a suitable $\underline{Q}$. As an
example consider the algebra generated by $\1$ alone. It can easily
be shown that any $\star$-sub-algebra is of this form if and only if
it contains a maximal abelian sub-algebra. Also from 
$E_{\underline{Q}}({\cal B})\;\underline{Q}$ may be recovered. Indeed
the $Q_{j}$'s are just the minimal self-adjoint idempotents (i.e. the
orthogonal projections) in $E_{\underline{Q}}({\cal B})$ and which are
central. Also
on the set of all spectral resolutions of the identity we introduce a partial
ordering $\le$ by setting $\underline{P}\le \underline{Q}$ if to each $i$
there is $j(i)$ (which is unique) such that $P_{i}\le Q_{j(i)}$. Note
that each $j$ is of the form $j=j(i)$ for at least one $i$. 
Then in particular all $P_{i}$ commute with all $Q_{j}$. 
Also $\underline{P}\le \{\1\}$ holds for all $\underline{P}$.
It is easy to see that $\underline{P}\le \underline{Q}$ if and only if 
$E_{\underline{P}}({\cal B})\subseteq E_{\underline{Q}}({\cal B})$.
With respect to these orderings $\underline{P}$ or equivalently 
$E_{\underline{P}}({\cal B})$ is minimal
 if and only if each $P_{i}$ is one-dimensional. 
$E_{\underline{P}}({\cal B})$ is then commutative with dimension equal
to $\dim{\cal H}$. To sum up, with respect to the partial ordering
$\le$ there is a unique maximal element but there are many minimal
elements in the set of spectral resolutions $\underline{P}$.
 
Now one has
the well known result $S(E_{\underline{Q}}(\rho))\ge S(\rho)$ 
(see e.g. \cite{Nielsen} for a direct proof and \cite{Wehrl} for
the special case when $\dim Q_{j}=1$ for all $j$. It is a special case of 
 Uhlmann's monotonicity theorem \cite{U2}, see also \cite{OP}). It
 means that projective measurements increase entropy
and compares with the inequality $S(\rho|\sigma)\le S(\rho)$ to
be proven below. Its interpretation is that of a projective 
measurement described by
the family $\underline{Q}$ of projections on a system given by $\rho$, 
but where we never learn of the result of the
measurement. In contrast $S(\rho|\sigma)$ is interpreted as
a set of projective measurements given by the projections $Q_{j}$, 
each performed with the probability $\dim\, Q_{j}\;\sigma_{j}$, and 
where we learn of each 
outcome $F(\rho,Q_{j})$ separately. The sum in \eqref{def1} and 
\eqref{def3} then reflects the 
occurrence of a quantum decoherence. In other words one 
considers the family of density
operators $\rho_{Q_{j}},\; Q_{j}\rho Q_{j}\neq 0$, takes their von
Neumannn entropy and then forms the linear combination with the
non-negative coefficients $\tr Q_{j}\sigma\;\tr Q_{j}\rho$.

By definition we have
\begin{equation}
\label{F1}
F(\rho,Q=\1)=S(\rho|\sigma=(1/\dim\1)\cdot\1)=S(\rho).
\end{equation}
We consider this property to be necessary for any
other sensible definition of a conditional entropy involving two
density matrices. It holds for Shannon's 
conditional entropy $H(X|Y)$ in the form $H(X|Y)=H(X)$ when $Y$ is the
trivial partition (see section 3), which
means that there is no gain in information, if $Y$ contains no
information. We will return to this point in section 3.

Some additional remarks are in order. Since the quantity $S(\rho|\sigma)$ is 
supposed to be a quantum mechanical mechanical analogue of Shannon's
conditional entropy $H(X,Y)$, $\rho$ corresponds to $X$ and $\sigma$
to $Y$. In analogy to the classical case, where $X$ and $Y$ may be
considered to be stochastic variables living on the same space, here
the density matrices $\rho$ and $\sigma$ also live on the same space.
Unfortunately with this correspondence $S(\rho|\sigma)$ does not   
reduce to the classical case when $\rho$ and $\sigma$ commute (see
\eqref{cond4} and its discussion in section 3). As matter of fact, we 
shall argue in section 4 that a quantum conditional entropy with this 
property does not exist. 

By construction we have the obvious invariance under
unitary automorphisms
\begin{equation}
\label{U}
F(\rho,Q)=F(U\rho U^{-1},UQU^{-1}),
\end{equation}
for any $U\in\cal{U}(\cal{H})$, the group of unitary operators in
$\cal{H}$.
This relation \eqref{U} immediately implies 
\begin{equation}
\label{U1}
S(U\rho U^{-1}|U\sigma U^{-1})=S(\rho|\sigma)
\end{equation}
for all $U$. Relation \eqref{U1} reflects the fact that $S(\rho|\sigma)$ is
defined intrinsically and is in particular basis independent.
Therefore this invariance property should also hold for any
alternative, sensible definition of a quantum mechanical conditional
entropy defined in terms of two density matrices.
We shall comment on the classical analogue to \eqref{U1} in section 4.

The next observation is also important. It is easy to see that 
$F(\rho,Q)$ is continuous in $\rho$ and $Q$ by the same arguments used to prove
continuity of $S(\rho)$. Therefore $S(\rho|\sigma)$ is also
continuous in $\rho$ for fixed $\sigma$. However, $S(\rho|\sigma)$ is not  
continuous in $\sigma$ everywhere for all fixed $\rho$. It is
continuous on the dense open subset
where the eigenvalues of $\sigma$ are non-degenerate.In fact, it is
zero there(see below). So this lack of
continuity occurs where $\sigma$ has degenerate eigenvalues and is due to
the fact that for $Q=Q^{\prime}+Q^{\prime\prime}$ being the sum 
of two projections both $\neq 0$ and which are orthogonal to each other, i.e. 
$Q^{\prime}Q^{\prime\prime}=0$,  
in general one has
\begin{equation}
\label{disc}
\dim\,Q\,F(\rho,Q)\neq \dim\,Q^{\prime}\,F(\rho,Q^{\prime})
+\dim\,Q^{\prime\prime}\,F(\rho,Q^{\prime\prime}).
\end{equation}
 To understand this consider the case when $\dim {\cal
  H}=2$. Then $S(\rho|\sigma)=S(\rho)$ if $\sigma =1/2\;\1$ and 
$S(\rho|\sigma)=0$ otherwise.
At the moment we do not know whether this lack of continuity of 
$S(\rho|\sigma)$ in $\sigma$ is
a desirable feature or not, i.e whether this can be understood
quantum mechanically, when we interpret $S(\rho|\sigma)$ as the entropy
of $\rho$ conditioned by $\sigma$. 
Observe that a degeneracy typically 
occurs when a non-trivial symmetry is present. In other words there
is then a 
non-trivial non-abelian subgroup ${\cal G}={\cal G}(\sigma)$ of ${\cal U}({\cal H})$ such
that $U\sigma U^{-1}=\sigma$ for all $U\in{\cal G}$. Note that 
${\cal G}$ always contains a subgroup isomorphic to the abelian group 
$U(N=\dim\,{\cal H})$.
In this picture a removal of degeneracies is related to a breakdown of 
symmetry, a familiar phenomenon in physics. 

To proceed further, $F(\rho,Q)=0$ if $Q\rho Q=0$, which can
happen for $Q\neq 0$ only if $\rho$ has zero as an eigenvalue, i.e. if
$\rho$ is not strictly positive. Then
also $(\1-Q)\rho Q=Q\rho(\1-Q)=0$. In fact, by Schwarz inequality 
for any $\psi,\psi^{\prime}\in {\cal H}$ we have  
\begin{equation}
|<\psi,Q\rho(\1-Q)\psi^{\prime}>|\le ||\rho^{1/2}Q\psi||||\rho^{1/2}
(\1-Q)\psi^{\prime}||=0.\nonumber
\end{equation}
This also shows that $Q\rho Q=0$ is equivalent to $Q\rho =0$, which
 in turn by the self-adjointness of $\rho$ and $Q$ is equivalent to 
$\rho Q=0$. By the trivial identity
\begin{equation}
\label{rho}
\rho =Q\rho Q+(\1-Q)\rho Q+Q\rho(\1-Q)+(\1-Q)\rho(\1-Q),
\end{equation}
valid for all $\rho,Q$,
we therefore also have $\rho=(\1-Q)\rho(\1-Q)$ whenever
$Q\rho Q=0$. Obviously \eqref{rho} gives 
$\tr\rho=\tr Q\rho Q+\tr(\1-Q)\rho(\1-Q)$  such that in
particular the inequalities $0\le \tr Q\rho Q\le 1$ and 
$0\le \tr(\1-Q)\rho(\1-Q)\le 1$ hold for any $\rho$ and $Q$. By relation \eqref{F} we also have
$F(\rho,Q)\ge 0$ and hence $S(\rho|\sigma)\ge 0$ for all $\rho,Q$ and 
$\sigma$.
Now $S(\rho_{Q})=0,Q\rho Q\neq 0$ holds if and only if $\rho_{Q}$ is
a pure state, i.e. a one-dimensional projection. Also for $\dim Q=1$
one always has $Q\rho Q=(\tr Q\rho Q)Q$. We collect this observation
in
\begin{lem}
$F(\rho,Q)=0$ if and only if $Q\rho Q$ is a multiple of a
one-dimensional projection.
\end{lem}
This multiple is allowed to be zero.
To characterize such $Q$'s fulfilling the conditions of the lemma, let 
$P(\rho)\neq 0$ be the projection operator onto the subspace corresponding
to the non-zero eigenvalues, such that $P(\rho)\rho=\rho=\rho P(\rho)$
and in particular
$P(\rho)=\1$ if $\rho>0$. Using the spectral representation of $\rho$
it is easy to see that $Q\rho Q$ is a
multiple (possibly zero) of a one-dimensional projection if and only 
if $Q$ may be
written as $Q=Q^{\prime}+Q^{\prime\prime}$ with $\dim Q^{\prime}\le 1$
and $P(\rho)Q^{\prime\prime}=\rho Q^{\prime\prime}=0$.

More generally  consider the case where $Q\rho Q=(\tr(Q\rho Q)/\dim
Q^{\prime})\cdot Q^{\prime},Q\neq 0$ holds for a suitable projection operator 
$Q^{\prime}$ such that in particular
$0\neq Q^{\prime}\le Q$ and $Q^{\prime}$ is unique whenever 
$Q\rho Q\neq 0$. Then 
$F(\rho,Q)=(\tr Q\rho Q)\ln\dim Q^{\prime}$ and 
$\rho_{Q}=(1/\dim Q^{\prime})Q^{\prime}$. 
This gives the 
\begin{lem} 
\label{le2}
If all non-zero eigenvalues of $\sigma$ are non-degenerate
then $S(\rho|\sigma)=0$ for all $\rho$. More generally if
$Q_{j}\rho Q_{j}$ is a multiple (possibly zero) of some projection operator 
$Q^{\prime}_{j}\,(\le Q_{j})$ 
for all $j$ with $\sigma_{j}>0$, then   
\begin{equation}
\label{Q1}
  S(\rho|\sigma)=\sum_{j}\tr Q_{j}\rho \;\tr Q_{j}\sigma
  \ln\dim Q^{\prime}_{j}.
\end{equation}
\end{lem}
Observe that $S(\rho|\sigma)=0$ for all pure states $\sigma$ and all $\rho$.
If $\rho$ is pure
then $Q\rho Q$ is always a multiple of a pure state for all
$Q$. Therefore $S(\rho|\sigma)=0$ also holds for all $\sigma$ whenever
$\rho$ is pure. 
Also if $\rho\sigma=0$
which is equivalent to $\tr\rho\sigma=0$ and
which can happen only if neither $\rho$ nor $\sigma$ is strictly
positive, then again $S(\rho|\sigma)=0$. 
Sufficient (but not necessary) for the condition of Lemma \ref{le2} to
hold is that to each $j$ with $\sigma_{j}>0$ there is $i(j)$ with 
$Q_{j}\le P_{i(j)}$. 
For these $j$'s $Q_{j}^{\prime}=Q_{j},\,Q_{j}\rho Q_{j}=\rho_{i(j)} Q_{j}$ and
hence $\tr Q_{j}\rho =\rho_{i(j)}\dim Q_{j}$. 
This gives in particular
\begin{equation}
\label{S2}
S(\rho|\rho)=\sum_{i}\rho_{i}^{2}(\dim P_{i})^{2}\ln\dim P_{i}.
\end{equation}
Therefore the relation $S(\rho|\rho)=0$  holds if and only if all the non-zero
eigenvalues of $\rho$ are non-degenerate, the if part being a special
case of Lemma \ref{le2}. 
  
If in addition to the property $Q_{j}\le P_{i(j)}$
 the density matrix 
$\sigma$ is such that 
\begin{equation*}
\sum_{j:\,i(j)=i}\sigma_{j}(\dim Q_{j})^{2}\ln\dim Q_{j}\le
\rho_{i}(\dim P_{i})^{2}\ln\dim P_{i}
\end{equation*}
 holds for all $i$, then by \eqref{Q1} and \eqref{S2} 
$S(\rho|\sigma)\le S(\rho|\rho)$.
Note that this last condition is satisfied if
\begin{equation*}
\sum_{j:\,i(j)=i}\sigma_{j}\,\dim Q_{j}\le
\rho_{i}\,\dim P_{i}
\end{equation*} 
holds since trivially $\dim\, Q_{j}\le\dim\, P_{i(j)}$. 

We return to a discussion of the general properties of $F(\rho,Q)$ and
$S(\rho|\sigma)$.
The first main result of this article shows that $S(\rho|\sigma)$
shares an important property with $S(\rho)$ (see
e.g. \cite{Lieb,Wehrl} for the classical and the quantum entropy and 
\cite{McEliece} for Shannon's conditional
entropy and derived quantities).
\begin{theo}
$F(\rho,Q)$ and $S(\rho|\sigma)$ are both concave in $\rho$.
\end{theo}
Again we consider this property to be necessary for any sensible
definition of a quantum conditional entropy. Like for the entropy
$S(\rho)$ itself it states that mixing (in $\rho$) increases
(conditional) entropy.
On the other hand the case $\dim {\cal H}=2$ discussed
above shows that in general  $S(\rho|\sigma)$ for fixed $\rho$ is
neither convex nor concave in $\sigma$. Intuitively it would be 
desirable to have concavity with respect to $\sigma$ since mixing 
the conditioning should increase conditional entropy.

The proof follows easily from the presentation \eqref{F2} and
\eqref{def4}
and the known convexity property of the relative entropy.  

The second main result of this article shows in particular that 
$S(\rho|\sigma)$ satisfies Shannon's inequality.
\begin{theo} The following inequalities hold for all density matrices 
$\rho$ and $\sigma$ in a fixed finite dimensional Hilbert space
\begin{equation}
\label{S3}
0\le S(\rho|\sigma)\le S(\rho).
\end{equation}
If $\rho>0$ the last inequality is strict unless $\sigma=(1/\dim\;\1)\cdot\1$. 
\end{theo}
The above comparison of $S(\rho|\sigma)$ with
$S(E_{\underline{Q}}(\rho)$ suggests
another definition of conditional entropy with the conditioning not
given in terms of a density matrix $\sigma$ but rather only in terms 
of any resolution $\underline{Q}$ of the identity.
\begin{equation}
\label{Snew}
S(\rho|\underline{Q})=\sum_{j}\frac{\dim\,Q_{j}}{\dim\,\1}F(\rho,Q_{j}).
\end{equation}
By \eqref{FS} below we have
\begin{equation*}
0\le S(\rho|\underline{Q})\le S(\rho),
\end{equation*}
where the first inequality is an equality if $\dim\,Q_{j}=1$ for all
$j$ and the second one an equality if the spectral resolution is
trivial, i.e. if 
$\underline{Q}=\{\1\}$. We note that in \eqref{Snew} any sequence of
numbers $\sigma_{j}^{\prime}\ge 0$ ( labeled in the same way as the 
$Q_{j}$'s) with
$\sum_{j}\sigma_{j}^{\prime}=1$ and replacing $\dim\,Q_{j}/\dim\,\1$
would do equally well. But then we may combine and encode these data
$\underline{Q}$ and $\{\sigma\}$ in the density matrix 
$\sigma=\sum_{j}\sigma_{j}\,Q_{j}$ with 
$\sigma_{j}=\sigma_{j}^{\prime}/\dim\,Q_{j}$. If in 
addition all the $\sigma_{j}$'s are pairwise different, then by our 
discussion above 
they and the spectral resolution $\underline{Q}$ may be recovered from 
$\sigma$ and we are back to our construction $S(\rho|\sigma)$.

Due to the relation $1=\tr \sigma=\sum_{j}\dim Q_{j}\;\sigma_{j}$ this
second theorem is an immediate consequence of the following
\begin{lem} 
\label{le3}
For all $\rho$ and $Q$ the inequality
\begin{equation}
\label{FS}
F(\rho,Q)\le S(\rho) 
\end{equation}
holds. If $\rho>0$ this inequality is strict unless $Q=\1$. 
\end{lem}
Before we turn to a proof we make some remarks.
We conjecture that in the general case $\rho\ge 0$, the inequality
\eqref{FS} is strict unless $Q\rho=\rho$. This would imply that the
second inequality in \eqref{S3} is strict unless 
$\sigma\rho=(\tr\sigma\rho)\rho$, which means the following. Any 
$\sigma$ with $\sigma\rho=(\tr\sigma\rho)\rho$ is of the form
$\sigma=(\tr\sigma\rho) P(\rho)+\sigma^{\prime}$ with 
$(\1-P(\rho))\sigma^{\prime}=\sigma^{\prime}$. 

Instead of 
$F(\rho,Q)$ one might be tempted to consider instead the quantity 
(see \eqref{def2})
\begin{equation*}
\tilde{F}(\rho,Q)=-\tr(Q\rho Q\ln Q\rho Q)\ge 0
\end{equation*}
and try to prove $\tilde{F}(\rho,Q)\le S(\rho)$. Obviously we have
$\tilde{F}(\rho,Q)\ge F(\rho,Q)$. Consider, however, the case where $\dim
Q=1$ and $\rho=P,\;\dim P=1$ (i.e. $\rho$ is pure) and with $P$ chosen such that $QPQ=(\tr QP)Q$ 
satisfies $0<\tr PQ <1$. Then
$0=S(\rho=P)<\tilde{F}(\rho =P,Q)$. Furthermore one has 
$F(\rho,Q)\le S(\rho_{Q})$ when $0<\tr Q\rho Q(\le 1)$. But it does
not make sense to replace $F(\rho,Q)$ by $S(\rho_{Q})$ as an alternative, since
$S(\rho_{Q})$ is only defined when $Q\rho Q\neq 0$. Even if $Q\rho
Q\neq 0$, one does not have $S(\rho_{Q})\le S(\rho)$ in general.
To see this we will consider an example. For any 
$0\neq\psi\in {\cal H}$ let $P_{\psi}$ be the  1-dim. projection onto
the subspace spanned by $\psi$.
\begin{ex}
Let $\dim\,{\cal H}=4$ with $\psi_{1},\psi_{2},\psi_{3},\psi_{4}$
being an orthonormal basis. Let $Q=P_{\psi_{1}}+P_{\psi_{2}}$ be the 2-dim. 
projection onto the sub-space spanned by $\psi_{1}$ and $\psi_{2}$.
Choose
$\rho(\phi_{1},\phi_{2})=\rho_{1}P_{\psi^{\prime}_{1}}
+\rho_{2}P_{\psi^{\prime}_{2}},\,
\rho_{1}+\rho_{2}=1$
with 
\begin{eqnarray*}
\psi^{\prime}_{1}&=&\cos\phi_{1}\,\psi_{1}+\sin\phi_{1}\,\psi_{3},\\
\psi^{\prime}_{2}&=&\cos\phi_{2}\,\psi_{2}+\sin\phi_{2}\,\psi_{4},\;
\cos\phi_{1}\neq 0\neq\cos\phi_{2}. 
\end{eqnarray*}
Then 
\begin{equation*}
\rho(\phi_{1},\phi_{2})_{Q}=\frac{\cos^{2}\phi_{1}\,\rho_{1}}
 {\cos^{2}\phi_{1}\,\rho_{1}+\cos^{2}\phi_{2}\,\rho_{2}}P_{\psi_{1}}+
         \frac{\cos^{2}\phi_{2}\,\rho_{2}}
 {\cos^{2}\phi_{1}\,\rho_{1}+\cos^{2}\phi_{2}\,\rho_{2}}P_{\psi_{2}}.          \end{equation*}
Assume $0<\rho_{1}<1$ such that $S(\rho(\phi_{1},\phi_{2}))\neq 0$
 and choose
 $\phi_{1}$ and $\phi_{2}$ such that 
$cos^{2}\phi_{1}\,\rho_{1}=\cos^{2}\phi_{2}\,\rho_{2}$. This gives
$\rho(\phi_{1},\phi_{2})_{Q}=1/2\,Q$ with 
$S(\rho(\phi_{1},\phi_{2})_{Q})=\ln 2>S(\rho(\phi_{1},\phi_{2}))$ 
whenever $\rho_{1}\neq 1/2$. On the other hand, some easy
estimates show that indeed 
$F(\rho(\phi_{1},\phi_{2}),Q)\le S(\rho(\phi_{1},\phi_{2}))$ holds for
all $\phi_{1}$ and $\phi_{2}$. 
\end{ex}
This example also shows that in general neither $\rho_{Q}$ nor
$S(\rho_{Q})$ for $Q\rho Q=0$ may be defined by a limiting
procedure. In fact, we may let $\phi_{1}$ and $\phi_{2}$ tend to
$\pi/2$ in such a way that $\cos^{2}\phi_{2}/\cos^{2}\phi_{1}$ tends
to an arbitrary constant $\ge 0$ showing that in the limit for 
$\rho(\phi_{1},\phi_{2})_{Q}$ we may
obtain an arbitrary convex combination of $P_{\psi_{1}}$ and
$P_{\psi_{2}}$ and hence an arbitrary value
between 0 and $\ln 2$ for the entropy. 
By the convexity of the relative entropy we also have
\begin{equation*} 
F(\rho,Q)+F(\rho,\1 -Q)\le S(E_{\{Q,\1-Q\}}(\rho)).
\end{equation*}
On the other hand, in general $F(\rho,Q)+F(\rho,\1-Q)$ is in general
not bounded above by $S(\rho)$. Indeed, consider the following
\begin{ex}
Let the set-up be as in Example 2.1. With repsect to this basis let
\begin{equation*}
\rho(\kappa)=\frac{1}{4}\left(\begin{array}{cccc}
1&0&0&\kappa\\
0&1&\kappa&0\\
0&\kappa&1&0\\
\kappa&0&0&1
\end{array}\right)
\end{equation*}
with $0\le\kappa\le 1$. The two two-fold degenerate eigenvalues are 
$1/4(1\pm\kappa)$. This gives $F(\rho,Q)+F(\rho,\1-Q)=\ln 2$ whereas
$S(\rho(\kappa))=\ln 2 -1/2((1+\kappa)\ln
(1+\kappa)+(1-\kappa)\ln(1-\kappa))<\ln 2$, whenever $0<\kappa$.
\end{ex}
The quantity 
\begin{equation}
\label{delta}
\Delta S(\rho)=S(\rho)-S(\rho|\rho)\ge -\sum_{i}\dim
P_{i}\;\rho_{i}\ln(\dim
P_{i}\;\rho_{i})
\end{equation}
is of special interest. The inequality is a consequence of $\dim
P_{i}\;\rho_{i}\le 1$ and again implies that the right hand side is 
non-negative and equal to zero
if and only if $\rho=(1/\dim\;\1)\cdot\1$ such that $\Delta S(\rho)>0$
unless $\rho=(1/\dim\;\1)\1$. In more detail the inequality in
\eqref{delta} may also be written as follows. Let 
$S_{cl}(\underline{p})\ge 0$ be the
classical entropy for the probability distribution 
$\underline{p}=(p_{1},p_{2},... p_{n}),p_{k}\ge 0,\sum_{k}p_{k}=1$ 
\begin{equation*}
S_{cl}(\underline{p})=-\sum_{k=1}^{n}p_{k}\;\ln p_{k}.
\end{equation*}
such that in particular
\begin{equation}
\label{Scl0}
S(\rho)=S_{cl}(\underline{p}(\rho))
\end{equation}
with
\begin{equation}
\label{prho}
\underline{p}(\rho)=(\underbrace{\rho_{1},..,\rho_{1}}_{\dim P_{1}},
                     \underbrace{\rho_{2},..,\rho_{2}}_{\dim P_{2}},....).
\end{equation}
\eqref{delta} may now be rewritten as 
\begin{equation}
\label{Scl}
0\le S_{cl}(\underline{\hat{p}}(\rho))\le \Delta S(\rho)
\end{equation}
with
\begin{equation*}
\underline{\hat{p}}(\rho)=(\dim P_{1}\;\rho_{1},\dim
P_{2}\;\rho_{2},...)
\end{equation*} 
and where $S_{cl}(\underline{\hat{p}})=0$ if and only if $\rho$ is a
pure state.
$\Delta S(\rho)$ is
easily shown to be continuous in $\rho$ and is obviously bounded above
by 
$\ln\dim\;\1=\ln\dim {\cal H}=S(\rho=(1/\dim\;\1)\cdot\1)$. 
It would be interesting to find its maximum in $\rho$ for
fixed dimension of $\dim\cal{H}$. Note also that
\begin{equation}
\label{S4}
S(\rho)=S_{cl}(\underline{p}(\rho))=S_{cl}(\underline{\hat{p}}(\rho))+
\sum_{i}\dim P_{i}\;\sigma_{i}\ln\dim P_{i}
\ge S_{cl}(\underline{\hat{p}}(\rho))
\end{equation} 
with equality if and only if $\dim P_{i}=1$ for all $i$ with
$\sigma_{i}>0$. We will discuss $\Delta S(\rho)$ below when
we compare $S(\rho|\sigma)$ with Shannon's conditional entropy. 

We turn to the proof of \eqref{FS}. First recall that $F(\rho,Q)$
is continuous in $\rho$ (and $Q$). Hence it suffices to consider the
case $\rho> 0$ which implies that $Q\rho Q \neq 0$ for all $Q\neq 0$. Since 
$F(\rho,Q)=0$ for $Q=0$ and $\dim Q=1$ and since $F(\rho,\1)=S(\rho)$ 
it suffices to consider the case $1<\dim Q<\dim\;\1$. 

Now $\cal{U}(\cal{H})$ operates transitively and continuously on the
Grassmannian of all $n$-dimensional subspaces of 
$\cal{H}$, $(1\le n\le \dim\cal{H})$.
For each $n$ this space is therefore compact and homeomorphic to the
set of all projections of dimension $n$. Obviously on this
set $\cal{U}(\cal{H})$ operates, again continuously, via $U: P\mapsto 
UPU^{-1}$. By \eqref{U} 
\begin{equation}
F_{n}(\rho)=\sup_{Q:\dim Q=n}F(\rho,Q)
=\sup_{U:U\in\cal{U}(\cal{H})}F(\rho,UQ_{0}U^{-1})
=\sup_{U:U\in\cal{U}(\cal{H})}F(U\rho U^{-1},Q_{0}),
\end{equation}
which is finite for each $n$. Here $Q_{0}$ is any orthogonal projection with
$\dim Q_{0}=n$. In particular we may choose $Q_{0}$ such that 
$F_{n}(\rho)=F(\rho,Q_{0})$. Consider the one-parameter unitary subgroup
$U(t)=\exp(-itK)$, where $K$ is an arbitrary self-adjoint operator on
$\cal{H}$. Then we must have $f_{K}(t)=F(U(t)\rho U(-t),Q_{0})\le
F(\rho,Q_{0})=f_{K}(t=0)$ for all $t$ and all s.a. $K$.
Now it is well known that for any one parameter family of strictly
positive operators $A(t)$ which is differentiable in $t$  one has
\begin{equation*}
\frac{d}{dt}\tr(A(t)\ln A(t))=\tr((\1+\ln A(t))\frac{d}{dt}A(t)).
\end{equation*}
Recalling the assumption $\rho > 0$ such that $Q_{0}\rho Q_{0}>0$ when
restricted to the subspace $Q_{0}\cal{H}$, it is easy to see that
$f_{K}(t)$ is also differentiable in $t$ at $t=0$ and
\begin{equation}
\label{deriv}
\begin{array}{ccc}
\frac{d}{dt}f_{K}(t)|_{t=0}&=&-i\;\tr((\1+\ln Q_{0}\rho
                         Q_{0})Q_{0}[K,\rho]Q_{0})\\
                  && +i\;(1+\ln \tr(Q_{0}\rho Q_{0}))\tr Q_{0}[K,\rho]Q_{0}\\
                  &=&i\;\tr([\rho,\ln \tr(Q_{0}\rho Q_{0})\cdot Q_{0} 
                             -Q_{0}(\ln Q_{0}\rho Q_{0})Q_{0}]K). 
\end{array}
\end{equation}
By definition of $Q_{0}$ we must have $d/dt f_{K}(t=0)=0$ for all $K$.
But then \eqref{deriv} implies that $\rho$ commutes with
$B=Q_{0}B=BQ_{0}$ given as
\begin{equation*}
B=\ln \tr(Q_{0}\rho Q_{0})\cdot Q_{0} 
                             -Q_{0}(\ln Q_{0}\rho Q_{0})Q_{0}.
\end{equation*}
This in turn implies that $\rho$ commutes with $Q_{0}$ itself, which is
easy to see. Indeed, use the spectral representation 
$Q_{0}\rho Q_{0}=\sum_{k}\rho^{\prime}_{k}Q^{\prime}_{k}$ with
$Q^{\prime}_{k}\le Q_{0},\dim Q^{\prime}_{k}=1$ and 
$\sum_{k}Q^{\prime}_{k}=Q_{0}$ to write $B$ as
\begin{equation*}
B=\sum_{k}(\ln (\sum_{l}\rho^{\prime}_{l})-\ln\rho^{\prime}_{k})Q^{\prime}_{k}.
\end{equation*}
Now write any $\psi\in Q_{0}\cal{H}$ as $\psi =\sum_{k}a_{k}\psi_{k}$,
where $\psi_{k}$ is a unit vector in $Q^{\prime}_{k}\cal{H}$. 
Set 
\begin{equation*}
\phi=\sum_{k}\frac{a_{k}}{(\ln
  (\sum_{l}\rho^{\prime}_{l})-\ln\rho^{\prime}_{k})}\psi_{k}\in
  Q_{0}{\cal H}.
\end{equation*}
$\phi$ is well defined since 
$\sum_{l}\rho^{\prime}_{l}\neq \rho^{\prime}_{k}$ for every
$k$. This follows from our assumption $n>1$, the fact that $\ln x$ is
strictly monotonic in $x$ and that $\rho^{\prime}_{k}>0$ for all $k$, since 
$Q_{0}\rho Q_{0}$ when restricted to $Q_{0}\cal{H}$ is strictly positive.
By construction $\psi=B\phi$ such that 
$\rho\psi=\rho B\phi=B\rho\phi=Q_{0}B\rho\phi\in Q_{0}\cal{H}$. Thus
$\rho$ leaves $Q_{0}\cal{H}$ invariant and hence commutes with
$Q_{0}$, as was claimed. But then we have 
$\rho=Q_{0}\rho Q_{0}+(\1-Q_{0})\rho(\1-Q_{0})$ which implies
\begin{equation*}  
S(\rho) =-\tr(Q_{0}\rho Q_{0}\ln Q_{0}\rho Q_{0})
         -\tr((\1-Q_{0})\rho(\1-Q_{0})\ln(\1- Q_{0})\rho(\1- Q_{0})). 
\end{equation*}
This gives
\begin{eqnarray}
\label{final}
S(\rho)&=F(\rho,Q_{0})&
        -\tr((\1-Q_{0})\rho(\1-Q_{0})\ln(\1- Q_{0})\rho(\1- Q_{0}))\nonumber\\
       &&-\tr Q_{0}\rho Q_{0}\ln \tr Q_{0}\rho Q_{0}.
\end{eqnarray}
The two last terms in \eqref{final}, however, are non-negative. This
concludes the proof of the claim \eqref{FS}. To prove the second part
of Lemma \ref{le3},
we observe that the last two terms in \eqref{final} vanish exactly when 
$(\1-Q_{0})\rho(\1-Q_{0})=0$. But this contradicts the assumption
$\rho>0$ and $\dim Q_{0}<\dim\;\1$, the case $Q=\1$ having been
discussed previously. This completes the proof of Lemma \ref{le3}.

\section{Comparison with the classical case}

In this section we provide a comparison with the classical theory of
Shannon (see \cite{Shannon} and for expositions
e.g. \cite{Ga,McEliece,Top}). 
For the convenience of the reader and in order to establish
notation we recall the basic facts. Let $\{\Omega,\mu\}$ be a
probability space. Furthermore let $X=\{X_{\alpha}\}$ and
  $Y=\{Y_{\beta}\}$ be any two partitions (up to measure zero) of 
$\Omega$ into disjoint subsets of non-zero measure. For simplicity we
will assume these partitions to be finite, i.e. we choose the
indices $\alpha$ and $\beta$ to be in the range
$1\le\alpha\le n, 1\le\beta\le m$.
Set $\underline{p}(X)=\{p_{\alpha}\}$ with $p_{\alpha}=\mu(X_{\alpha})>0$
and $\underline{p}(Y)=\{q_{\beta}\}$ with $q_{\beta}=\mu(Y_{\beta})>0$ 
such that $\sum_{\alpha}p_{\alpha}=1$ and
$\sum_{\beta}q_{\beta}=1$. Here and in what follows $\alpha$ is an
index referring to $X$ and $\beta$ to $Y$.
Then $H(X)=-\sum_{\alpha}p_{\alpha}\ln p_{\alpha}\ge 0$ and
similarly $H(Y)=-\sum_{\beta}q_{\beta}\ln q_{\beta}\ge 0$ is Shannon's
entropy. Actually Shannon used $\log _{2}$ instead of $\ln$ adapting
to the situation where information is coded in bits, but this
is not relevant for our purpose. Since $H(X)=S_{cl}(\underline{p}(X))$ 
this concept of information
theory relates to the concept of entropy in classical statistical mechanics.
Shannon's conditional entropy is now given as follows. Let
\begin{equation*}
p_{\alpha|\beta}=\frac{\mu(X_{\alpha}\cap Y_{\beta})}{\mu(Y_{\beta})},\qquad
q_{\beta|\alpha}=\frac{\mu(Y_{\beta}\cap X_{\alpha})}{\mu(X_{\alpha})}
\end{equation*}
be conditional probabilities associated to $X$ and $Y$ (i.e. 
$p_{\alpha|\beta}$ is the probability that $X_{\alpha}$ 
will happen, given that $Y_{\beta}$ has happened).
Obviously 
\begin{equation}
\label{Bayes}
p_{\alpha|\beta}\,q_{\beta}=q_{\beta|\alpha}\,p_{\alpha}
(=\mu(A_{\alpha}\cap B_{\beta}))
\end{equation} 
for all $\alpha,\beta$, which is called Bayes rule for
$p_{\alpha|\beta}$ and $q_{\beta|\alpha}$. 
Let $\underline{p}_{\beta}=(p_{1|\beta}, p_{2|\beta},..,p_{n|\beta})$ and 
$\underline{q}_{\alpha}=(q_{1|\alpha},
q_{2|\alpha},.... q_{m|\alpha})$, 
such that
\begin{equation}
\label{p1} 
\underline{p}=\sum_{\beta=1}^{m}q_{\beta}\underline{p}_{\beta},\qquad 
\underline{q}=\sum_{\alpha=1}^{n}p_{\alpha}\underline{q}_{\alpha}.
\end{equation}
Shannon's conditional entropy is now defined as
\begin{equation}
H(X|Y)=\sum_{\beta=1}^{m}q_{\beta}S_{cl}(\underline{p}_{\beta})
\end{equation}
and it satisfies
\begin{equation}
\label{H1}
0\le H(X|Y)\le H(X).
\end{equation}
We observe that the second inequality, called Shannon's inequality, is a 
consequence of the concavity of
the function $\underline{p}\mapsto S_{cl}(\underline{p})$ and
\eqref{p1}. It states that on average information on $X$ is gained
if $Y$ is known.
Also $0\le H(X,Y)=H(Y)+H(X|Y)$ is symmetric in $X$ and $Y$ and
satisfies
\begin{equation}
\label{H2}
H(Y)\le H(X,Y)\le H(X)+H(Y).
\end{equation}
Actually $H(X,Y)=H(X\vee Y)$, where $\vee$ denotes the join of two partitions.
The inequalities in \eqref{H1} and \eqref{H2} turn into equalities if 
the following 
conditions hold. $X$ and $Y$ are said to be independent if
$p_{\alpha|\beta}=p_{\alpha}$ holds for all $\alpha$ and
$\beta$. This means that
$\underline{p}_{\beta}$ is actually independent of $\beta$ and equals
$\underline{p}$ and
$\underline{q}_{\alpha}$ is independent of $\alpha$ and equals
$\underline{q}$. In particular  
$S_{cl}(\underline{p}_{\beta})=S_{cl}(\underline{p})$ holds for all $\beta$
and $S_{cl}(\underline{q}_{\alpha})=S_{cl}(\underline{q})$ for all $\alpha$.
The second inequality in \eqref{H1} and the second inequality in
\eqref{H2} (which are equivalent) are now equalities if and only if
$X$ and $Y$ are independent. It follows from the fact that 
$S_{cl}(\underline{p})$ is strictly concave in $\underline{p}$.
Secondly $X$ is called a consequence of $Y$ if to each $\alpha$ there is
$\beta(\alpha)$ such that $p_{\alpha|\beta(\alpha)}=1$. So this
means that $p_{\alpha|\beta}=0$ for all
$\beta\neq \beta(\alpha)$ and hence $S_{cl}(\underline{p}_{\beta})=0$ for all
$\beta$. Therefore the first inequality in
\eqref{H1} and equivalently the first inequality in \eqref{H2} are
equalities if and only if $X$ is a consequence of $Y$. In particular
\begin{equation}
\label{H3}
H(X|X)=0,
\end{equation}
i.e. $H(X,X)=H(X)$. 

With this brief review of Shannon's theory we turn to a comparison
with our quantum mechanical construction. Obviously \eqref{H1}
corresponds to \eqref{S3} when we let $X$
correspond to $\rho$ and $Y$ to $\sigma$.
Note, however, the difference between \eqref{H3} and \eqref{S2}.
Moreover for the quantity $S(\rho,\sigma)=S(\sigma)+S(\rho|\sigma)$ 
we have the inequalities
\begin{equation}
\label{S5}
S(\sigma) \le S(\rho,\sigma)\le S(\rho)+S(\sigma),
\end{equation}
which correspond to \eqref{H2}.
$S(\rho,\sigma)$ is in general not symmetric in
$\rho$ and $\sigma$ . To see this consider commuting $\rho$ and 
$\sigma$. Then we have
\begin{equation}
\label{cond4}
S(\rho|\sigma)=-\sum_{j,i}\dim\;Q_{j}\,\sigma_{j}\,\rho_{i}\,\dim(P_{i}Q_{j})
      \ln\frac{\rho_{i}}{\tr(\rho Q_{j})}.                                    
\end{equation}
We remark that if 
$\tr(\rho Q_{j})=0$ for a fixed $j$ then $\tr(P_{i}Q_{j})=0$ for all $i$. Also
\eqref{S2} is a special case of \eqref{cond4}. \eqref{cond4} shows 
that even in the commutative case $S(\rho,\sigma)$ is not 
symmetric in $\rho$ and $\sigma$. 
So this implies that in the commutative case
$S(\rho|\sigma)$ does not reduce to $H(X|Y)$ for {\em any} 
choice of $X=X(\rho)$ and $Y=Y(\sigma)$ with $H(X)=S(\rho)$
 and $H(Y)=S(\sigma)$. 
This lack of symmetry of $S(\rho,\sigma)$
is in contrast to the symmetry of its classical counterpart $H(X,Y)$, 
which has an important interpretation. The relation $H(X,Y)=H(Y,X)$ is 
equivalent to $H(Y)+H(X|Y)=H(X)+H(Y|X)$, a consequence of Bayes rule. 
But this means that on average 
the information on $Y$ plus the information on $X$
given $Y$ is equal to the information on $X$ plus the information on 
$Y$ given $X$. It would be interesting to see whether this failure of 
symmetry for $S(\rho,\sigma)$ has a sensible interpretation in the context of
the familiar {\em Alice and Bob} set-up in quantum information
theory, see e.g. \cite{Preskill}. 

Finally consider
\begin{equation}
\label{S6}
0\le
S(\rho||\sigma)=S(\rho)+S(\sigma)-S(\rho,\sigma)=S(\rho)-S(\rho|\sigma)\le
 S(\rho)
\end{equation}
which corresponds to
\begin{equation*}
0\le I(X||Y)=H(X)+H(Y)-H(X,Y)=H(X)-H(X|Y).
\end{equation*}
On average $0\le I(X||Y)\le H(X)$ gives the information 
gain on $X$ when knowing $Y$. Thus if there is no information content at all in
$Y$, i.e. if $Y$ is the trivial partition $\{\Omega\}$, then there is no 
information gain in $X$
\begin{equation}
\label{H4}
I(X||Y=\{\Omega\})=0.
\end{equation}
Thus \eqref{H4} corresponds to \eqref{F1} when rewritten as 
$S(\rho||\sigma=(1/\dim\,\1)\1)=0$.  
 Therefore we also interpret the quantum mechanical analogue 
$S(\rho||\sigma)$ as a quantum information gain for $\rho$ given
$\sigma$ and which by \eqref{S6} can be at most $S(\rho)$. In
particular the gain is maximal for all $\rho$, if all non-zero
eigenvalues of $\sigma$ 
are non-degenerate. The gain is also maximal if $\rho\sigma =0$, since
then $S(\rho|\sigma)=0$, see Lemma \ref{le2} and the remark thereafter.

Finally $\Delta S(\rho)$ (see \eqref{delta}) corresponds to $I(X||X)$
and describes the situation where $\rho$ is conditioned on itself,
$\sigma=\rho$. Then by \eqref{Scl} there is non-zero information gain
unless $\rho$ is pure (and then a gain is not necessary). In contrast
to the classical situation, $I(X||X)=H(X)$, which gives complete
information gain when $X$ is conditioned on itself, there
is complete information gain in the quantum case,
$\Delta S(\rho)=S(\rho)$, if and only if all 
non-zero eigenvalues of $\sigma$ are non-degenerate.

\section{Attempts of alternative constructions}

We conclude by addressing the natural question whether there is a quantity
$S^{?}(\rho|\sigma)$ which shares more properties with Shannon's
conditional entropy than the $S(\rho|\sigma)$ we have given.
More precisely and by the arguments given in the preceding sections 
it would be desirable for $S^{?}(\rho|\sigma)$ to have
(most of) the following properties
\begin{enumerate}

\item Invariance under the group ${\cal U}({\cal H})$:
   $S^{?}(U\rho U^{-1}|U\sigma U^{-1})=S^{?}(\rho|\sigma)$ for all 
   $U\in{\cal U}({\cal H})$ (compare \eqref{U1}).
\item Bounds: $0\le S^{?}(\rho|\sigma)\le S(\rho)$ for all $\rho$ and
   $\sigma$ with $S^{?}(\rho|\rho)=0$ and \\
   ${S^{?}(\rho|\sigma=(1/\dim\;\1)\1)=S(\rho)}$.
\item Classical equivalence with Shannon's conditional entropy.
\item Symmetry: $S^{?}(\rho,\sigma)=S(\sigma)+S^{?}(\rho|\sigma)$ is 
 symmetric in $\rho$ and $\sigma$.
\item Continuity of $S^{?}(\rho|\sigma)$ in $\rho$ and in $\sigma$.
\item Concavity of $S^{?}(\rho|\sigma)$ in
      $\rho$ and $\sigma$.
\end{enumerate}
Note that $S(\rho|\sigma)$ fulfills condition 1, condition 2 apart
from the property $S(\rho|\rho)=0$, condition 5 up to a set of measure
zero and condition 6 only with respect to $\rho$.  

Both the equality requirements of condition 2 can never be satisfied 
simultaneously. Indeed,
with the choice $\rho=\sigma=1/\dim {\cal H}\,\1$ we should have both
$S(1/\dim {\cal H}\,\1|1/dim{\cal H}\,\1)\\
=0$ and 
$S(1/\dim {\cal H}\,\1|1/dim{\cal H}\,\1)=\ln\dim{\cal H}$. Also the
condition $S(\rho|\rho)=0$ combined with $S(\rho|\sigma)\ge 0$ is 
incompatible with concavity of 
$S(\rho|\sigma)$ in $\rho$ (condition 6). In fact, let 
$\rho=\lambda\, \rho_{1}+(1-\lambda)\,\rho_{2},\,0<\lambda<1$. But this gives
$0=S(\rho|\rho)\ge\lambda\,S(\rho_{1}|\rho)+(1-\lambda)S(\rho_{2}|\rho)$. 
Hence $S(\rho_{1}|\rho)=0$ for all $\rho_{1}$ for which there is
$\lambda>0$ with $\lambda\,\rho_{1}<\rho$. This condition is
fulfilled for all $\rho_{1}$, whenever $\rho>0$ (I owe these observations
to H. Narnhofer).

Next let us look at the condition 3, by which we mean the situation
where $\rho$ and $\sigma$ commute such that  
$S(\rho)=H(X),\,S(\sigma)=H(Y)$ and 
$S^{?}(\rho|\sigma)=H(X|Y)$ holds for suitable $X=X(\rho)$ and
$Y=Y(\sigma)$. Also the dependence of $X(\rho)$ and $Y(\sigma)$ on
$\rho$ and $\sigma$ respectively should be non-trivial w.r.t. their 
eigenvalues. In particular condition 3 means that the symmetry condition 4
must hold at least when $\rho$ and $\sigma$ commute. 
In view of the destruction of quantum coherence when measurements 
are performed and due to the occurrence of the sum by which 
$S^{?}(\rho,\sigma)$ is defined, it is unclear to the author whether 
the symmetry condition 4 also should hold for non-commuting $\rho$ and
$\sigma$ (see below , however, a construction of conditional entropy 
in terms of spectral resolutions of the identity below). It is natural 
to make the assumption on $X(\rho)$, that $\mu(X_{k})= p_{k}(\rho)$,
see \eqref{Scl0}. Then it may be shown that the 
continuity condition and the classical equivalence condition are not 
compatible. The concavity
condition in $\sigma$ is at least intuitively desirable since taking
convex combinations decreases conditioning, i.e. increases uncertainty,
and hence should increase conditional entropy.

We would also like to point out another 
difference between the classical and the quantum case in the way we
have presented it so far. In the classical case
the conditioning $Y$ is trivial when $Y=\{\Omega\}$, which means no 
information content and for which we have $H(Y)=0$. Within the context of
density matrices the only sensible candidate for a trivial conditioning is
$\sigma=1/\dim{\cal H}\,\1$, since this is the density matrix with no
information content. Its von Neumann entropy, however, is
maximal. Recall that we used this quantum notion of trivial
conditioning in our discussion of the inequality $S(\rho|\sigma)\le
S(\rho)$ (see also the discussion following \eqref{H4}).
We note that several authors
consider von Neumann's entropy not to be a good
generalization of classical entropy (see e.g. \cite{Be}, page 141). 
In fact, in classical theory finer partitions give rise to higher
uncertainty and hence to larger classical entropy. This was the reason
for the algebraic approach of Connes and
St{\o}rmer and of Connes, Narnhofer and Thirring, in which a classical finer
partitioning corresponds to a larger algebra. In particular the larger
the algebra, the larger the
entropy and similarly the larger the conditioning algebra the larger
the conditional entropy. 

We claim, however, that there is a way to
reconcile this with von Neumann's entropy. Indeed, given a quantum
system in the state $\rho$, the measurements 
one can perform without disturbing $\rho$ are given by the observables
(i.e. the self-adjoint operators) in ${\cal A}(\rho)$, which by definition
is the  $\star$-sub-algebra of ${\cal B}$ consisting of all elements
in ${\cal B}$ which commute with $\rho$. In particular 
${\cal A}(\rho=1/dim{\cal H}\,\1)={\cal B}$. In this sense again
larger uncertainties 
correspond to larger algebras. In other words, the larger the entropy
the more measurements on can perform without disturbing the system in
the given state $\rho$. 
To be more precise, we introduce a
partial ordering $\preceq$ on the set of
all density matrices (which differs from the one introduced by
Uhlmann, see e.g. \cite{Wehrl}).
By definition $\rho\preceq\sigma$ ($\sigma$ is more mixed than
$\rho$), if and only if a) $\underline{P}\le\underline{Q}$ and b) 
$\tr\rho\,Q_{j}=\tr\sigma\,Q_{j}=\sigma_{j}\tr Q_{j}$ holds for all
$j$. It is easy to see that $\rho\preceq\sigma$ and
$\sigma\preceq\tau$ implies $\rho\preceq\tau$ and that 
$\rho\preceq 1/\dim{\cal H}\,\1$ and $\rho\preceq\rho$ holds for all
$\rho$. So whenever 
$\rho\preceq\sigma$ then condition a) implies 
${\cal A}(\rho)\subseteq {\cal A}(\sigma)$ and a) and b) combined
imply $S(\rho)\le S(\sigma)$ by the concavity of the von Neumann entropy.
Note, however, that the correspondence between $\rho$ and 
${\cal A}(\rho)$ is not one-to-one. In fact, ${\cal A}(\rho)$ only
depends on the spectral resolution of the identity $\underline{P}=
\underline{P}(\rho)$ 
associated to $\rho$ and not on the eigenvalues $\rho_{i}$ of $\rho$.
Indeed, one has ${\cal A}(\rho)=E_{\underline{P}(\rho)}({\cal B})$, as
one may easily verify.

Returning to our discussion of conditions 1-6, there is a way out,
however, if one considers spectral resolutions
of the identity $\underline{P}$ instead of density matrices. It works as follows.
First observe that the actual choice of the probability space 
$\{\Omega, \mu\}$ for Shannon's theory is irrelevant. What is relevant
are the the sets of non-negative numbers
$\underline{p}=\{p_{\alpha}\},\\\underline{q}=\{q_{\beta}\},$
$\underline{p}\vee\underline{q}=\{p_{\alpha|\beta}\}$ and 
$\underline{q}\vee\underline{p}=\{q_{\beta|\alpha}\}$ subject to the 
following conditions of which the last one is Bayes rule
\begin{equation}
\label{3cond}
\sum_{\alpha} p_{\alpha}=\sum_{\beta}q_{\beta}=1,\;
\sum_{\beta} p_{\alpha|\beta}q_{\beta}=p_{\alpha},\;
\sum_{\alpha} q_{\beta|\alpha}p_{\alpha}=q_{\beta},\;
 p_{\alpha|\beta}\,q_{\beta}=q_{\beta|\alpha}\,p_{\alpha}. 
\end{equation}
Note that then
\begin{eqnarray*}
\sum_{\alpha}p_{\alpha|\beta}&=&
\frac{1}{q_{\beta}}\sum_{\alpha}q_{\beta|\alpha}p_{\alpha}=1\\
\sum_{\beta}q_{\beta|\alpha}&=&
\frac{1}{p_{\alpha}}\sum_{\beta}p_{\alpha|\beta}q_{\beta}=1.
\end{eqnarray*}
We consider these conditions \eqref{3cond}, which mean independence of
a particular realization of partitions $X$ and $Y$ on a probability space, the
classical analogue of the relation \eqref{U1}.
Setting $p_{\alpha,\beta}=p_{\alpha|\beta}q_{\beta}$ and 
$q_{\beta,\alpha}=q_{\beta|\alpha}p_{\alpha}$, Bayes rule gives 
$p_{\alpha,\beta}=q_{\beta,\alpha}$. We will therefore 
write $H(X|Y)=H(\underline{p}|\underline{q})$ by a slight abuse of
notation since all the data
$\underline{p},\,\underline{q},\,\underline{p}\vee\underline{q}$ and 
$\underline{q}\vee\underline{q}$
in \eqref{3cond} are necessary for a specification of $H(X|Y)$. But given
these data it makes sense to say
that  $\underline{p}$ is a consequence of $\underline{q}$ or that
$\underline{p}$ and $\underline{q}$ are independent.

Now let $\tau=1/\dim{\cal H}\tr$
denote the normalized trace, i.e. $\tau(\1)=1$. For any two
spectral resolutions
$\underline{P}$ and $\underline{Q}$ let 
$p_{i}=\tau(P_{i}),\,q_{j}=\tau(Q_{j}),\,p_{i|j}
=\tau(P_{i}Q_{j})/\tau(Q_{j}),\,q_{j|i}=\tau(Q_{j}P_{i})/\tau(P_{i})$.
Note that by definition all $P_{i}$ and all $Q_{j}$ are non-zero
projections. The conditions 
\eqref{3cond} are obviously satisfied. We then set
$H(\underline{P})=S_{cl}(\underline{p}),
H(\underline{Q})=S_{cl}(\underline{q})$ ,such that
$H(\underline{Q}=\{\1\})=\ln\dim{\cal H}$ and finally 
$H(\underline{P}|\underline{Q})=H(\underline{p}|\underline{q})$, such
that $0\le H(\underline{P}|\underline{Q})\le H(\underline{P})$ as
desired. Note that now $H(\underline{P}|\underline{Q})$ is completely
specified by $\underline{P}$ and $\underline{Q}$. Also
$H(\underline{P},\underline{Q})=
H(\underline{Q})+H(\underline{P}|\underline{Q})$
is symmetric in $\underline{P}$ and $\underline{Q}$.

It is easy to
see that $\underline{p}$ is a consequence of $\underline{q}$ if and
only if $\underline{Q}\le \underline{P}$ such that 
$H(\underline{P}|\underline{Q})=0$ if and only if $\underline{Q}\le
\underline{P}$. Similarly $\underline{p}$ and $\underline{q}$ are
independent if and only if $\underline{P}=\{\1\}$ or
$\underline{Q}=\{\1\}$. Therefore, whenever $H(\underline{P})\neq 0$,
$H(\underline{P}|\underline{Q})=H(\underline{P})$ if and only if 
$\underline{Q}=\{\1\}$, which in this context is the trivial
conditioning and for which the entropy is zero in contrast to our
construction in terms of density matrices. Finally we set 
$Ad\,U \underline{Q}=\{U\,Q_{i}\,U^{-1}\}$
for any $\underline{Q}$ and any unitary $U$. Then obviously
$H(Ad\,U\underline{P}|Ad\,U\underline{Q})=H(\underline{P}|\underline{Q})$ 
(compare condition 1). 

Since the $P_{i}$'s and the $Q_{j}$'s need not commute, this
construction is a non-commutative version of Shannon's
conditional entropy in (commutative) classical probability theory. Thus a
classical partition $X$ is replaced by a spectral resolution 
of the identity $\underline{P}$, which in turn corresponds to the
$\star$-algebra $E_{\underline{P}}({\cal B})$ and which is abelian if
and only if each $P_{i}$ is one-dimensional. The choice $\underline{Q}=\{\1\}$ 
giving maximal entropy $H(\underline{Q})$ and maximal
conditional entropy $H(\underline{P}|\underline{Q})$ corresponds to the 
maximal algebra $E_{\underline{Q}=\{\1\}}({\cal B})={\cal B}$. Our 
construction of $H(\underline{P}|\underline{Q})$ differs from the 
construction in \cite{CS,CNT}.

We might have defined the conditional entropy of two density
matrices $\rho$ and $\sigma$ by
$H(\underline{P}(\rho)|\underline{Q}(\sigma))$. Conditions 1,2 and 4 are
then satisfied but not condition 5 and condition 3, since the dependence on
the eigenvalues of $\rho$ and $\sigma$ drops out. We conjecture that condition
6 is also not satisfied.

\textbf{Acknowledgements:} The author would like to thank M. Karowski,
 H. Narnhofer, M.A. Nielsen, M. Schmidt and E. St{\o}rmer for helpful remarks.


\markright{References}

\end{document}